\documentclass[twocolumn]{aastex61}
\usepackage{graphicx,color}
\usepackage{times}
\usepackage{amssymb}
\usepackage{xcolor}
\usepackage{amsmath}

\newcommand{\bs}{\boldsymbol}
\newcommand{\mrm}{\mathrm}

\newcommand{\hy}{\hat{\boldsymbol{y}}}
\newcommand{\hz}{\hat{\boldsymbol{z}}}
\newcommand{\dux}{\delta u_x}
\newcommand{\duy}{\delta u_y}
\newcommand{\duz}{\delta u_z}
\newcommand{\dbx}{\delta b_x}
\newcommand{\dby}{\delta b_y}
\newcommand{\dbz}{\delta b_z}
\newcommand{\etah}{\eta_{\mathrm{H}}}
\newcommand{\tetah}{\tilde{\eta}_{\mathrm{H}}}
\newcommand{\ua}{v_{\mathrm{A}}}

\newcommand{\etao}{\eta_{\mathrm{O}}}
\newcommand{\etaa}{\eta_{\mathrm{A}}}
\newcommand{\etap}{\eta_{\mathrm{P}}}
\newcommand{\etat}{\eta_{\mathrm{T}}}
\newcommand{\tetap}{\tilde{\eta}_{\mathrm{P}}}
\newcommand{\sigp}{\sigma_{\mathrm{P}}}
\newcommand{\sigv}{\sigma_{\nu}}
\newcommand{\sigm}{\sigma_{\alpha}}
\newcommand{\Pm}{\mathrm{Pm}}
\newcommand{\omga}{\omega_{\mathrm{A}}}
\newcommand{\omgh}{\omega_{\mathrm{H}}}
\newcommand{\omgp}{\omega_{\mathrm{P}}}
\newcommand{\omgv}{\omega_{\mathrm{\nu}}}
\newcommand{\omgg}{\omega_{\mathrm{G}}}

\newcommand{\kappah}{\kappa_{\mathrm{H}}}
\newcommand{\kc}{\tilde{k}_{\mathrm{c}}}
\newcommand{\km}{\tilde{k}_{\mathrm{m}}}
\newcommand{\ki}{\tilde{k}_{\mathrm{i}}}
\newcommand{\kn}{k_{\mathrm{n}}}
\newcommand{\gammam}{\gamma_{\mathrm{m}}}
\newcommand{\ej}{e_{j}}
\newcommand{\eg}{e_{\gamma}}

\newcommand{\pder}[2]{\frac{\partial#1}{\partial#2}}
\newcommand{\pdder}[2]{\frac{\partial^2#1}{\partial#2^2}}

\begin{document}

\title{Spectral Analysis of Non-Ideal MRI Modes: The effect of Hall diffusion
}
\shorttitle{Non-Ideal MRI Modes}
\author{Gopakumar~Mohandas \& Martin~E.~Pessah}
\shortauthors{Mohandas \& Pessah}
\affiliation{Niels Bohr International Academy, Niels Bohr Institute, Blegdamsvej 17, DK-2100, Copenhagen~\O, Denmark}
\email{gopakumar@nbi.ku.dk, mpessah@nbi.ku.dk}

\begin{abstract}
The effect of magnetic field diffusion on the stability of
accretion disks is a problem that has attracted
considerable interest of late. In particular, the Hall effect has the
potential to bring about remarkable changes in the dynamical behavior of disks
that are without parallel.
In this paper, we conduct a systematic examination of the linear eigenmodes in
a weakly magnetized differentially rotating gas with special focus on
Hall diffusion.
We first develop a geometrical representation
of the eigenmodes and provide a detailed quantitative description of the
polarization properties of the oscillatory modes under the combined
influence of the Coriolis and Hall effects. We also analyze the effects of
magnetic diffusion on the structure of the unstable modes and
derive analytical expressions for the kinetic
and magnetic stresses and energy densities associated with the non-ideal MRI.
Our analysis explicitly demonstrates that, if the dissipative
effects are relatively weak, the kinetic stresses and energies make up the
dominant contribution to the total stress and energy density when the equilibrium
angular momentum and magnetic field vectors are anti-parallel. This is in sharp
contrast to what is observed in the case of the ideal or dissipative MRI.
We conduct shearing box
simulations and find very good agreement with the results derived from
linear analysis. As the modes in consideration are also exact solutions of
the non-linear equations, the unconventional nature of the kinetic and
magnetic stresses may have significant implications for the non-linear
evolution in some regions of protoplanetary disks.
\end{abstract}

\keywords{magnetohydrodynamics ---  instabilities --- accretion disks }

\section{Introduction}
\label{sec:introduction}

The magnetorotational instability (MRI, \citealt{bh98}), driven by
differential rotation and weak magnetic fields, is
considered to be the foremost mechanism of linear destabilization in
astrophysical disk systems.
There has been substantial ongoing interest in studying the effect of
magnetic field diffusion on the MRI primarily with a view to understanding
protoplanetary disk evolution \citep{turneral14}.
In particular, diffusion mediated by Hall currents has commanded a great
deal of attention by virtue of its capacity to
pave the way to new avenues of destabilization \citep{w99,bt01}.
Local linear analysis has helped reveal the
markedly different character of the unstable dynamics \citep{w99,bt01,ws12}
and their fundamental dependence on disk conditions, namely, the
relative orientation of the net equilibrium angular momentum and magnetic
field vectors and the strength of the Hall currents.

One expects to find vast swathes within a protoplanetary disk
that are conducive to the prevalence of significant Hall currents as a result
of ion-neutral collisions \citep{kb04,pw08,armitage11}. This has provided
great impetus in driving efforts to understand the non-linear evolution
of disks influenced by non-ideal effects. A number of local
shearing box simulations with Hall diffusion either in
isolation or in unison with other non-ideal effects (viz. ohmic and ambipolar
diffusion) have been carried out in the recent past
\citep{ss02a,ss02b,bejarano2011,kl13,lesuretal2014,bai2014,bai2015,simon15}.
Efforts are currently underway to perform global simulations
including the Hall effect and the first among them has already been reported
by \citet{bethune16}.

While the march to conduct ever more sophisticated numerical experiments of a
non-ideal MHD disk system strides onwards, certain fundamental aspects,
especially those pertaining to the question of angular momentum transport may
be beneficially served by a systematic examination of the
non-ideal MRI eigenmodes.
With this goal in mind, we revisit the local linear analysis of a uniformly
magnetized disk with Hall diffusion in the shearing sheet approximation. We
adopt the approach of \citet{pcp06a,pc08} that has previously been employed
to thoroughly examine the ideal and dissipative MRI eigenmodes. Here, we carry
out an exhaustive analysis of the detailed eigenmode structure of the unstable
and oscillatory modes affected primarily by Hall diffusion.
As part of our analysis, we determine the mean kinetic and magnetic
stresses and energy densities of the non-ideal MRI mode
across parameter space.
Our work reveals that the relative dominance of the mean Reynolds and
Maxwell stresses as well as the ratio of magnetic to kinetic energy can
deviate from that of ideal or dissipative MRI when the background field and
angular momentum vector are anti-parallel. These departures depend intimately
on the range of length scales involved and may have significant implications
for the ensuing turbulence.
A detailed analysis of the linear eigenmodes may
also find utility in testing and benchmarking numerical
algorithms designed to include Hall diffusion.

This paper is organized as follows. In Section \ref{sec:basics}, we outline the
fundamental assumptions and equations involved.
In Section \ref {sec:eigenproblem}, we layout the basic groundwork for
our analysis and solve the eigenvalue problem.
We then examine the mode properties in detail and
provide a physical picture of mode behaviour in Section \ref{sec:eigenmodes}.
In Section \ref{sec:energetics}, we discuss the properties of the kinetic
and magnetic stresses and energy densities for the unstable mode.
We present the results of numerical simulations in Section \ref{sec:simulations}
to test the validity of our analytical results and conclude with a summary and
discussion of the potential implications in Section \ref{sec:discussion}.

\section{Basic Equations and Assumptions}
\label{sec:basics}

We consider a partially ionized, weakly magnetized,
incompressible  gas subject to ohmic, Hall and
ambipolar diffusion in
the presence of a gravitational field due to a central point mass.
While we shall strive to retain generality wherever possible, our
primary focus will nevertheless be on characterizing the effect of
Hall diffusion on the linear modes.

We work in the shearing sheet \citep{golb65b} approximation and therefore
adopt a frame of reference that co-rotates at a fiducial radius,
$r_0$, in the midplane of the disk. The shearing sheet frame is defined by the
set of cartesian coordinates
\[
x = r-r_0, \quad y = r(\phi - \Omega_0 t), \quad z = z,
\]
where $x/r_0 \sim \varepsilon \ll 1$
and is based on a local expansion of the combined gravitational and
centrifugal potentials to first order in $\varepsilon$ around the fiducial
radius. The angular frequency at the fiducial radius is denoted by
$\Omega_0$ and the disk is assumed to be in dominant centrifugal balance
with the
radial gravitational force. Consequently, all other dynamical state variables
are taken to be uniform to lowest order in $\varepsilon$.
Ignoring vertical stratification, the incompressible shearing sheet equations
are given by
\begin{align}
\label{eq:ss-momentum}
  \pder{\bs{u}}{t} + (\bs{u} \cdot \nabla)\bs{u} &=
  2\bs{u}\times \bs{\Omega}_0 + q\Omega_0^2\nabla x^2 \nonumber \\
  & \phantom{=} -\frac{1}{\rho}\nabla\left(P + \frac{B^2}{8 \pi} \right)
  + \frac{(\bs{B} \cdot \nabla)\bs{B}}{4 \pi \rho} +
  \nu \nabla^2 \bs{u},  \\
\label{eq:ss-induction}
 \pder{\bs{B}}{t} &= \nabla \times (\bs{u}_e \times \bs{B})
 -\frac{c}{\sigma} \nabla \times \bs{J}
 , \\
\label{eq:ss-incompressible}
 \nabla \cdot \bs{u} &= 0, \\
\label{eq:ss-gauss}
 \nabla \cdot \bs{B} &= 0,
\end{align}
where $\rho$ is the gas density, $P$ is the gas pressure, $\bs{B}$ is the
magnetic field, $\sigma$ is the constant electrical conductivity, $c$ is the
speed of light and $\nu$ is the constant fluid viscosity.
The shear rate $q$ evaluated at the fiducial radius is defined as
\[
q = \left. -\pder{\ln \Omega}{\ln r} \right|_{r=r_0}.
\]
Here, $\bs{u}$ is the velocity of the neutrals and the electron velocity,
$\boldsymbol{u}_e$, may be expressed as \citep{bt01}
\begin{align}
\label{eq:evelocity}
  \bs{u}_e &=
  \bs{u} + (\bs{u}_e -\bs{u}_i) + (\bs{u}_i - \bs{u})
  = \bs{u} -\frac{\bs{J}}{e n_e} + \frac{\bs{J} \times \bs{B}}{\gamma_d \rho
  \rho_i c},
\end{align}
where $e$ is the electron charge, $n_e$ is the electron number density,
$\gamma_d$ is the drag coefficient and $\rho_i$ is the ion mass density.
The current density is given by
\begin{align}
\label{eq:currentdensity}
  \bs{J} &= \frac{c}{4 \pi}(\nabla \times \bs{B}).
\end{align}

Equations (\ref{eq:ss-momentum})--(\ref{eq:ss-gauss}) admit
$\bs{u} = -q \Omega_0 x \hy$ and $\bs{B} = B_0 \hz$ as a steady-state
solution for the velocity and magnetic field\footnote{Note that Equations
(\ref{eq:ss-momentum})-(\ref{eq:ss-gauss})
are insensitive to the presence of a uniform background toroidal field under axial symmetry.}.
We consider Eulerian perturbations ($\delta \bs{u}, \delta \bs{B}$) to all the
fluid variables which are assumed to depend only on the vertical
coordinate and time.
Rescaling the Eulerian magnetic field perturbations to have dimensions
of velocity, $\delta \bs{b} \equiv \delta \bs{B}/\sqrt{4\pi\rho}$,
we obtain the following set of linearized equations
\begin{align}
\label{eq:lin-ux}
  \pder{\dux}{t} &= 2\Omega_0\duy +\ua \pder{\dbx}{z} + \nu \pdder{\dux}{z},
  \\
\label{eq:lin-uy}
  \pder{\duy}{t} &= (q-2)\Omega_0\dux + \ua \pder{\dby}{z}
  + \nu \pdder{\duy}{z} , \\
\label{eq:lin-bx}
  \pder{\dbx}{t} &=  \ua \pder{\dux}{z}
  + \frac{cB_0}{4\pi e n_e}\pdder{\dby}{z} \nonumber \\
  & \phantom{=} + \left( \frac{c^2}{4\pi\sigma} + \frac{B_0^2}{4\pi\rho\gamma
  \rho_i} \right) \pdder{\dbx}{z}, \\
\label{eq:lin-by}
  \pder{\dby}{t} &= \ua \pder{\duy}{z} -\frac{cB_0}{4\pi e n_e}\pdder{\dbx}{z}
  - q\Omega\dbx \nonumber \\
  & \phantom{=} + \left( \frac{c^2}{4\pi\sigma} +
  \frac{B_0^2}{4\pi\rho\gamma\rho_i} \right) \pdder{\dby}{z}.
\end{align}
We have also defined the equilibrium Alfv\'en speed as
\begin{equation}
\label{eq:uA}
  \ua \equiv \frac{B_0}{\sqrt{4 \pi \rho_0}}.
\end{equation}

The constraints of incompressibility, Equation (\ref{eq:ss-incompressible}),
and solenoidality, Equation (\ref{eq:ss-gauss}), require that $\duz = \dbz =$
const and we may thus set $\duz = \dbz = 0$ without loss of
generality.
Furthermore, restricting the spatial dependence of the perturbations to the
vertical dimension implies that non-linear terms vanish exactly
from Equations (\ref{eq:lin-ux})--(\ref{eq:lin-by}). Therefore, even though we
refer to the problem at hand as a linear mode analysis, the modes under
consideration are expected to be long-lived \citep{goxu94}.

\section{Eigenvalue Problem}
\label{sec:eigenproblem}

We conduct the linear analysis by solving the eigenvalue
problem defined in the shearing sheet frame.
The basic analysis in this setting has been carried out in a number
of previous studies
(\citealt{w99,bt01,kunz08,ws12}). We shall however, closely inspect
the characteristics of the linear eigenmodes that will enable us to
establish fundamental properties of the mean kinetic and magnetic stresses
and energy densities.

Assuming vertically periodic boundary conditions over the domain $[-H,H]$,
where $2H$ may be taken to be the vertical extent of the disk, we express the
perturbed variables as a Fourier series in $z$, such that
\begin{equation}
  \delta f(z, t) = \sum_{n=-\infty}^{\infty}  \hat{\delta f}(\kn, t)
  \exp(i
  \kn
  z),
\end{equation}
where $\kn = n\pi/H$, with $n$ an integer number and $\delta f$
represents any of the given Eulerian perturbations\footnote{For weak
magnetic fields,  we may approximate
$\Delta k = k_{\mathrm{n+1}} - \kn \propto \beta^{-1/2}$ and thus consider the
distribution of  wavenumbers to be approximately continuum even for moderate
values of the plasma $\beta\sim O(10^{2-3})$.}.
In what follows, we shall omit the subscript $\textrm{n}$ for the wavenumber as well as the
subscript $0$ for the equilibrium variables for brevity and convenience.

The set of Equations (\ref{eq:lin-ux})--(\ref{eq:lin-by}), can be expressed
more compactly as
\begin{equation}
  \label{eq:eigequation}
  \pder{}{t}\bs{\hat{\delta}}(k, t) = \bs{\mrm{L}}\bs{\hat{\delta}}(k, t),
\end{equation}
where
\begin{align}
  \label{eq:vectorhat}
  \bs{\hat{\delta}}(k, t) =
  [\hat{\dux} \quad \hat{\duy} \quad \hat{\dbx} \quad \hat{\dby}]^{\mrm
  {T}},
\end{align}
and the linear operator $\bs{\mrm{L}}$ is
\begin{align}
\label{eq:linearoperator}
  \bs{\mrm{L}} =
  \left[\begin{array}{cccc}
    -\omgv & 2\Omega & i\omga & 0 \\
    (q-2)\Omega & -\omgv & 0 & i\omga \\
    i\omga & 0 & -\omgp & -\omgh \\
    0 & i\omga &\omgh - q\Omega & -\omgp
  \end{array} \right] \, ,
\end{align}
which we have expressed entirely in terms of the frequencies defined below
\begin{align}
  \omga &\equiv k \ua, & \text{Alfv\'en frequency}  \\
  \omgv &\equiv k^2\nu, & \text{Viscous frequency}  \\
  \omgp &\equiv k^2\etap, & \text{Pedersen frequency}  \\
  \omgh &\equiv k^2\etah. & \text{Hall frequency}
\end{align}
Here we have also introduced the Pedersen diffusivity
\begin{equation}
  \etap = \etao + \etaa \equiv
  \frac{c^2}{4\pi\sigma} + \frac{B^2}{4\pi\rho\gamma_d\rho_i},
\end{equation}
with $\etao$ and $\etaa$ denoting the ohmic and ambipolar diffusivities
respectively, as well as the Hall diffusivity
\begin{equation}
\label{eq:etah}
  \etah \equiv \bs{\hat{\Omega}}\cdot\bs{\hat{B}}\frac{cB }{4\pi e n_e} =
  s|\etah|
  \,.
\end{equation}
The parameter $s$ assumes the value of $\pm1$ depending on
the value of the scalar product $\bs{\hat{\Omega}}\cdot\bs{\hat{B}}$
in Equation (\ref{eq:etah})\footnote{With more general wavevectors and angular
frequency profiles, the sign of $\etah$ is determined by
the quantity $(\bs{k}\cdot\bs{\omega})(\bs{k}\cdot \bs{B})$,
where $\bs{\omega} = \nabla \times \bs{u}$ is the equilibrium vorticity
\citep{kunz08}.}.

The linear operator $\bs{\mrm{L}}$ has four eigenvalues,
$\sigma_j$, and associated eigenvectors,
$\bs{e}_j$, that satisfies the eigenvalue equation
\begin{equation}
  \bs{\mrm{L}}\bs{e}_j =
  \sigma_j \bs{e}_j
  \quad \text{for }j = 1,\dots,4.
\end{equation}
$\bs{\mrm{L}}$ is a normal operator and therefore
its eigenvectors are orthogonal if the associated eigenvalues are
non-degenerate.
In this case, the eigenvectors of $\bs{\mrm{L}}$ constitute a linearly
independent basis set and thus any given arbitrary vector
$\hat{\bs{\delta}}$ can be represented as the linear
combination
\begin{equation}
\label{eq:delta-basisexp}
  \bs{\hat{\delta}} = \sum_{j = 1}^{4} a_j\bs{e}_j,
\end{equation}
where $a_j$ are in general complex valued time dependent quantities and
may be thought of as the coordinates in the $\mathbb{C}_4$ space defined by
the eigenvectors. Substituting Equation (\ref{eq:delta-basisexp}) in
Equation (\ref{eq:eigequation}), we obtain
\begin{equation}
  a_j(t) = a_j(0)e^{\sigma_j t}.
\end{equation}
Therefore
\begin{equation}
  \bs{\hat{\delta}}(k,t) = \sum_{j = 1}^{4} a_j(0)e^{\sigma_j t} \bs{e}_j.
\end{equation}

\subsection{Dispersion relation and eigenvalues}
\label{sec:eigenvalues}

The characteristic polynomial derived from the matrix operator
$\bs{\mrm{L}}$, given by Equation (\ref{eq:linearoperator}) yields the
dispersion relation
\begin{align}
  \label{eq:dispersion}
  (\sigv\sigp + \omga^2)^2 -2q\Omega^2(\sigp^2
  + \omga^2) + 4\Omega^2 \sigp^2 \nonumber \\ +
  (\sigv^2 + \kappa^2)\kappah^2 + (4-q)\Omega\omgh\omga^2 = 0.
\end{align}
where
\begin{equation}
\kappa = \sqrt{2(2-q)}\Omega \quad \text{and} \quad
\kappah = \sqrt{\omgh(\omgh-q\Omega)},
\end{equation}
are the epicyclic and the Hall-epicyclic frequency respectively.
Defining $\kappah$ makes it easier to recognize the parallel between the
Hall-Shear Instability \citep{rudkit05,kunz08} that occurs when $\kappah^2 <
0$ and the well-known Rayleigh instability that is present when $\kappa^2 <
0$. We also use the shorthands,
\begin{equation}
\label{eq:sigvsigp}
\sigv = \sigma + \omgv \quad \text{and} \quad
\sigp = \sigma + \omgp.
\end{equation}

The dispersion relation Equation (\ref{eq:dispersion}) is rather cumbersome to
solve analytically when dissipative effects are included.
Nevertheless, we sketch the procedure for obtaining the roots below.
We begin by converting Equation (\ref{eq:dispersion}) to \emph{depressed} form
\begin{equation}
\label{eq:depressed}
  \sigm^4 + L\sigm^2 + M\sigm + N = 0,
\end{equation}
with the coefficients
\begin{align}
  L &= 2(\omga^2 - \alpha^2) + \kappa^2 + \kappah^2, \\
  M &= -2\alpha(\kappa^2 - \kappah^2), \\
  N &= (\omga^2 - \alpha^2)^2 + \kappa^2(\omga^2 + \alpha^2) -
  4\omga^2\Omega^2  \nonumber \\
  &\phantom{=} + \kappah^2(\kappa^2 + \alpha^2) + (4-q)\Omega\omgh\omga^2,
\end{align}
where $\sigm = (\sigv + \sigp)/2$ and $\alpha = (\omgv - \omgp)/2$.

The solutions of Equation (\ref{eq:depressed}) are given by
\begin{equation}
\label{eq:sol1}
  \sigm = \pm_a \sqrt{-\Lambda \mp_b \sqrt{\Delta}} \pm_b \frac{M}{4\sqrt
  {\Delta}},
\end{equation}
with
\begin{equation}
\label{eq:lamdel}
  \Lambda = \frac{3L}{4} + \frac{y}{2}, \quad \text{ and } \quad
  \Delta = (y + L)^2 - N,
\end{equation}
where $a$ and $b$ in Equation (\ref{eq:sol1}) mark the four possible
combination of the $\pm$ signs and $y$ is the solution of the cubic equation
\begin{align}
\label{eq:yeqn}
  \left( y + \frac{L}{2} \right)[(y + L)^2 - N] = \frac{M^2}{8}.
\end{align}
Provided $y \neq -L/2$, we may recast Equation (\ref{eq:yeqn}) as
\begin{equation}
  \sqrt{(y + L)^2 - N} = \frac{M/4}{\sqrt{L/4 + y/2}},
\end{equation}
and substituting in Equation (\ref{eq:sol1}), we obtain
\begin{equation}
\label{eq:sol2}
  \sigm = \pm_a \sqrt{-\left( \frac{3L}{4} + \frac{y}{2} \pm_b
  \frac{M/4}{\sqrt{L/4 + y/2}} \right)}
  \pm_b \sqrt{\frac{L}{4} + \frac{y}{2}}.
\end{equation}
Finally, using the shorthands defined in Equation (\ref{eq:sigvsigp}), we
obtain the eigenvalues,
\begin{equation}
\label{eq:sigmaj}
  \sigma_j = \sigm - \frac{\omgv+ \omgp}{2} \quad \text{where}
  \quad j=1, 2, 3, 4.
\end{equation}
Two of the solutions given by Equation (\ref{eq:sigmaj}) are oscillatory and
two are exponentially varying.
We derive asymptotic expressions for the eigenvalues in the
dissipationless limit $\omgv = \omgp = 0$ in Appendix
\ref{sec:evasymptotics}.

\section{The Eigenmodes}
\label{sec:eigenmodes}

The set of normalized eigenvectors of the operator $\bs{\mrm{L}}$,
Equation (\ref{eq:linearoperator}), can be expressed as
\begin{equation}
  \label{eq:normeigvector}
  \bs{\hat{e}}_{j} = \frac{\bs{e}_j}{|\bs{e}_j|} \quad \text{for}
  \quad j=1, 2, 3, 4,
\end{equation}
where
\begin{align}
\label{eq:fulleigvector}
  \bs{e}_{j} = \left[
  \begin{array}{c}
    2\Omega\sigp^2 + 2\Omega\kappah^2 + \omgh\omga^2  \\
    \sigp(\sigp\sigv + \omga^2) + \sigv\kappah^2 \\
    i\omga(2\Omega\sigp - \omgh\sigv) \\
    i\omga(\sigp\sigv + \omga^2 - 2q\Omega^2 + 2\Omega\omgh)
  \end{array}  \right] \, .
\end{align}

The eigenvector components satisfy the following relationship
\begin{align}
\label{eq:evrelationship}
  -\frac{\ej^4}{\ej^1} = \frac{\ej^3}{\ej^2}
  + \frac{\omga q\Omega(\omgp-\omgv)}{\ej^2},
\end{align}
where the superscripts denote the corresponding eigenvector component.
In the absence of Hall diffusion $\omgh \to 0$,
Equation (\ref{eq:fulleigvector})
reduces to Equation (48) of \citet{pc08} and to Equation (32) of
\citet{pcp06a} in the ideal limit, $\omgv = \omgp = \omgh = 0$.

In the dissipationless limit but including Hall diffusion,
multiplying Equation (\ref{eq:fulleigvector}) with
\begin{equation}
  \frac{i}{\omga[2\Omega(q\Omega-\omgh) - (\sigma^2 + \omga^2)]} \,,
\end{equation}
and using the identity (derived from the dispersion relation)
\begin{equation}
\frac{2\Omega\sigma_j^2 + 2\Omega\kappah^2 + \omgh \omga^2}
{2\Omega(q\Omega-\omgh) - (\sigma_j^2 + \omga^2)}
= \frac{\sigma_j^2 + \omga^2 + \kappah^2}{2\Omega - \omgh} \,,
\end{equation}
we may recast Equation (\ref{eq:fulleigvector}) in the more useful form
\begin{align}
\label{eq:halleigvector}
  \bs{\hat{e}}_{j} = \left[ \frac{i F}{\omga} \quad
  \frac{i \sigma_j F}{\omga G} \quad
  -\frac{\sigma_j}{G} \quad
  1 \right]^{\mrm{T}} \,,
\end{align}
where
\begin{align}
  F &= (\sigma_j^2 + \omga^2 + \kappah^2)(2\Omega - \omgh)^{-1}  \,, \\
  G &= [2\Omega(q\Omega-\omgh) - (\sigma_j^2 + \omga^2)](2\Omega - \omgh)^{-1} \,.
\end{align}
The physically meaningful perturbation components are then obtained from the
real part of the eigenvector as
\begin{equation}
\label{eq:realdelta}
 \bs{\delta}_j(z,t) =  \Re[\bs{\hat{\delta}}(k,t) \exp(i k z)].
\end{equation}
Since $\bs{\delta_j}$ is a function of the real spatial variable $z$ and time
$t$, we can draw geometrical meaning from the eigenvector, Equation
(\ref{eq:realdelta}), and construct a physical picture of the mode evolution.

A defining property is the relative orientation of the velocity and
magnetic field components associated with the perturbations by taking the
scalar product of the two dimensional vectors defined by $\bs{\delta_u} =
[\ej^1 \ \ej^2]$ and $\bs{\delta_b} = [\ej^3 \ \ej^4]$,
i.e.,  $\bs{\delta_u} \cdot \bs{\delta_b} = u_0 b_0 \cos\theta_{j}$, where
\begin{equation}
  u_0= \sqrt{|\ej^1|^2 + |\ej^2|^2}   \text{ and }
    b_0= \sqrt{|\ej^3|^2 + |\ej^4|^2} \,
\end{equation}

In what follows, it shall be expedient, on occasion, to use the
dimensionless variables
\begin{equation}
\tilde{k} = \frac{k \ua}{\Omega}, \quad
\tetah = \frac{\etah\Omega}{\ua^2}, \quad
\tetap = \frac{\etap\Omega}{\ua^2} .
\end{equation}

\subsection{The Oscillatory Eigenmode}
\label{sec:oscillatoryem}

\begin{figure*}
  \centering
  \includegraphics{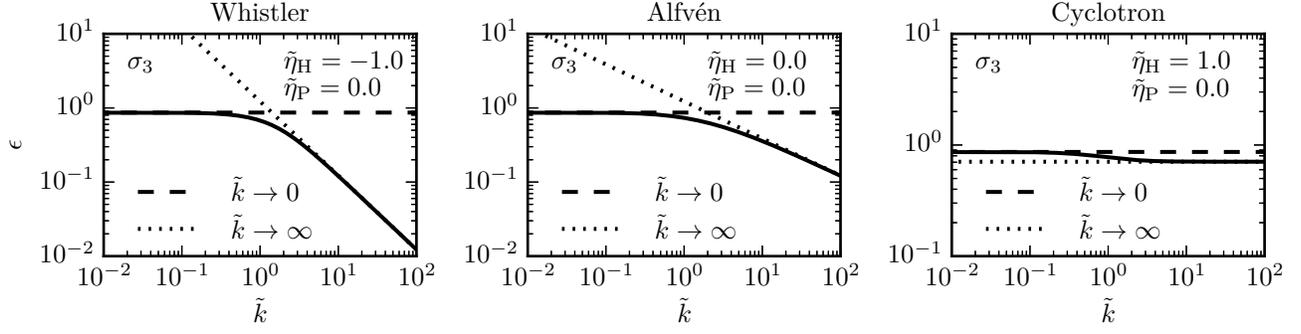}
  \caption{The eccentricity $\epsilon$ of the polarized oscillatory mode,
  $\sigma_3 = i
  \omega$, as a function of wavenumber for $q = 1.5$ and $\tetah = -1.0, 0.0,
  1.0$. Asymptotic approximations in the low wavenumber limit
  ($\tilde{k} \to 0$) and the high wavenumber limit ($\tilde{k} \to \infty$)
  are represented by the dashed and dotted lines respectively. The left panel
  corresponds to a Whistler mode, the central panel corresponds to an Alfv\'en
  mode and the right panel corresponds to a cyclotron mode; all three
  subject to the combined influence of rotation and shear (see Appendix
  \ref{sec:evasymptotics}).}
  \label{fig:hevp-exentricity}
\end{figure*}

The Hall effect is distinct from the other kinds of magnetic diffusion in
that the electromotive forces it induces act as a
``magnetic-Coriolis'' force \citep{bt01}.
This property leads to the polarization of the oscillatory eigenmodes in a
manner akin to that rendered by the kinetic Coriolis force. The only effect
that ohmic and ambipolar
diffusion has on these modes is to damp the wave amplitude over time.
Since the effect of dissipation on the eigenmodes has been studied
extensively in \citet{pc08}, we shall focus exclusively on the geometric
aspects of the oscillatory modes due to Hall diffusion alone and set $\omgv
= \omgp = 0$ here.

In order to provide a geometrical representation of the modes in physical space, it is useful to
first consider the norm of the ratios
\begin{equation}
\label{eq:ev-normratios0}
  \left| \frac{e_j^2}{e_j^1} \right|^2 = \left| \frac{e_j^3}{e_j^4} \right|^2
  = \frac{|\sigma_j|^2 (2\Omega-\omgh)^2}
  {\left[ 2\Omega(q\Omega - \omgh) - (\sigma_j^2 + \omga^2 )\right]^{2}}.
\end{equation}
Note that we retain the label $j$ to denote the eigenmode here
as the unstable modes may also become oscillatory beyond a cut-off
wavenumber for certain values of the Hall parameter.
Using the dispersion relation, Equation (\ref{eq:dispersion}), the ratio
defined in Equation (\ref{eq:ev-normratios0}) becomes
\begin{align}
\label{eq:ev-normratios}
  \left| \frac{e_j^2}{e_j^1} \right|^2 = \left| \frac{e_j^3}{e_j^4} \right|^2
  = -\frac{|\sigma_j|^2 \sigma_j^{-2}}{1 + \mu} ,
\end{align}
where we have defined the quantity
\begin{equation}
  \mu \equiv
  q \Omega \frac{[\sigma_j^2 + \omga^2 + 2\Omega(\omgh - q\Omega)]}
  {\sigma_j^2(2\Omega - \omgh)}.
\end{equation}
When the mode is purely oscillatory, $-|\sigma_j|^2\sigma_j^{-2} = 1$ and
Equation (\ref{eq:ev-normratios}) simply describes an ellipse with
the components of $\bs{\delta_u}$ and $\bs{\delta_b}$
representing the semi-major and minor axes.
The eccentricity of the ellipse, $\epsilon$, is related to $\mu$ as
\begin{align}
\label{eq:epsilon-mu}
  \epsilon^2 = \begin{cases}
  |\mu| & \quad \text{if }\mu<0, \\
  \mu/(1 + \mu) &\quad \text{if }\mu>0.
  \end{cases}
\end{align}
With the aid of the asymptotic forms for the eigenvalues,
Equations (\ref{eq:asym-k0})--(\ref{eq:asym-kinf}), we can determine the asymptotic
behaviour of the eccentricity as given below
\begin{align}
  -2 < \tetah & \leq -1/2: \nonumber \\
  \sigma_3:
  & \lim_{k \to 0} \epsilon^2 \sim \frac{q}{2} \, ,
  && \lim_{k \to \infty} \epsilon^2 \sim \frac{q \Omega}{|\omgh|} \, ,
  \\
  \nonumber \\
  -1/2 < \tetah & < 0: \nonumber \\
  \sigma_1:
  &
  && \lim_{k \to \infty} \epsilon^2 \sim \frac{q\Omega|\omgh|}
  {\omga^2+2\Omega|\omgh|} \, ,
  \nonumber \\
  \sigma_3:
  & \lim_{k \to 0} \epsilon^2 \sim \frac{q}{2} \, ,
  && \lim_{k \to \infty} \epsilon^2 \sim \frac{q\Omega}{|\omgh|} \, ,  \\
  \tetah=0: \nonumber \\
  \sigma_1:
  &
  && \lim_{k \to \infty} \epsilon^2 \sim \frac{q\Omega}{\omga} \, ,
  \nonumber \\
  \sigma_3:
  & \lim_{k \to 0} \epsilon^2 \sim \frac{q}{2} \, ,
  && \lim_{k \to \infty} \epsilon^2 \sim \frac{q\Omega}{\omga} \, ,  \\
  \nonumber \\
  \tetah>0: \nonumber \\
  \sigma_1:
  &
  && \lim_{k \to \infty} \epsilon^2 \sim \frac{q\Omega}{\omgh} \, ,
  \nonumber \\
  \sigma_3:
  & \lim_{k \to 0} \epsilon^2 \sim \frac{q}{2} \, ,
  && \lim_{k \to \infty} \epsilon^2 \sim \frac{q\Omega|\omgh|}
  {\omga^2+2\Omega\omgh} \, .
\end{align}
The eccentricity of the Alfv\'en and Whistler modes
(see Appendix \ref{sec:evasymptotics} for mode nomenclature) decreases with
increasing
wavenumber and the polarization becomes increasingly circular. The eccentricity is generally maximum in the limit $k \to 0$, and
has the value $\epsilon_{\mathrm{max}} = \sqrt{q/2}$, which incidentally
shares the value of the Oort constant for a differentially rotating disk.
The eccentricity of the cyclotron mode (see Appendix \ref{sec:evasymptotics})
is only marginally lower
than the maximum $\epsilon_{\mathrm{max}}$ at large wavenumbers as its frequency is
bounded at $\omgg$, see Appendix \ref{sec:evasymptotics}. In Figure \ref{fig:hevp-exentricity},
we show the three distinct ways in which the eccentricity of the oscillatory
mode can vary as a function of the wavenumber with the asymptotic forms
derived above to match.

\begin{figure*}
  \centering
  \includegraphics{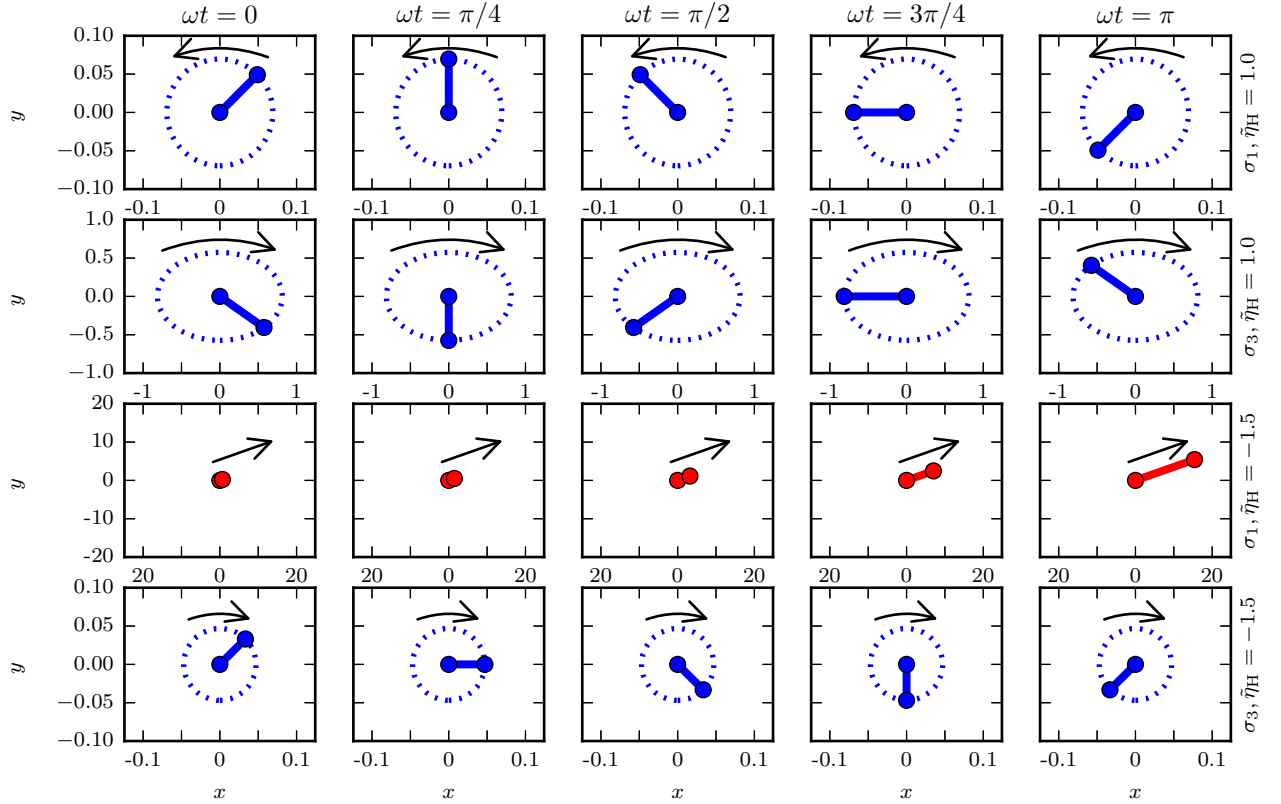}
  \caption{Visualization of the eigenmodes $\sigma_1$ and $\sigma_3$ in the
  horizontal plane at $kz= \pi/4$ for $\tilde{k} = 10, q =
  1.5$ and
  two different values of the Hall diffusivity, $\tetah = 1.0,
  -1.5$. The thick solid line denotes the velocity vector and the arrows
  indicate the direction of polarization (right or left, See Appendix
  \ref{sec:evasymptotics}) of the
  corresponding eigenmodes as seen by an observer looking down at the
  midplane from above. Each row of plots depicts the mode evolution over
  one half period in time.}
  \label{fig:hevp-geo}
\end{figure*}

Using Equations (\ref{eq:evrelationship}) and (\ref{eq:ev-normratios}), the
relative orientation of
$\bs{\delta_u}$ and $\bs{\delta_b}$ for the oscillatory modes can be
described by the angle
\begin{equation}
\label{eq:costhetaomega}
   \cos{\theta_{\omega}} = -\sqrt{\frac{1 - \epsilon^2(k)}
   {[1 - \epsilon^2(k)\cos^2 \varphi][1-\epsilon^2(k) \sin^2 \varphi]}} \, ,
\end{equation}
where $\varphi = k z + \omega t$.
In general, $\theta_{\omega}$ oscillates in time, so
$\bs{\delta_u}$ and $\bs{\delta_b}$ move in and out of
phase as $\varphi$ changes by a factor of $\pi/2$.

Figure \ref{fig:hevp-geo} charts the evolution of the net velocity vector
of the positive branch eigensolutions, $\sigma_1$ and $\sigma_3$,
over a half-period for a fixed wavenumber and two different values of the Hall
parameter. Notice that the polarization of $\sigma_1$ for
$\tetah = 1$ as well as $\sigma_3$ for $\tetah = -1.5$
is very nearly circular whereas the polarization of $\sigma_3$ for
$\tetah = 1$ is visibly elliptical. We also remind the reader that
any determination of the direction of polarization (right or left) is to be
made by examining the eigenvector, Equation (\ref{eq:realdelta}).
For instance, $\sigma_1$ associated with
$\tetah = 1$ is right elliptically polarized whereas $\sigma_3$ associated
with $\tetah = -1.5$ is left elliptically polarized even though both
behave like a Whistler mode at large wavenumbers.

\subsection{The Non-Ideal MRI Eigenmode}
\label{sec:unstableem}

Here, we examine the properties of the eigenvector corresponding to the
non-ideal MRI mode. Closed form expressions are much easily
derived in the absence of viscous effects and so we shall set $\omgv = 0$
hereafter.
This would correspond to considering the very low magnetic Prandtl number
limit $\mathrm{Pm} \equiv \nu/\etap \to 0$, which is also the relevant regime
of
parameter space with regard to protoplanetary disks.

We express below the main characteristic scales associated with the
unstable mode obtained from the dispersion relation,
Equation (\ref{eq:dispersion}) in
the inviscid limit \citep{ws12} and applicable in the parameter space
defined by $(\tetah, \tetap)$.

The critical wavenumber beyond which the non-ideal MRI is cut-off is
\begin{equation}
\label{eq:kc}
  \tilde{k}_{\mrm{c}}^2 = \frac{2q[1 + (2-q)\tetah]}
  {1 + (4-q)\tetah + 2(2-q)(\tetah^2 + \tetap^2)}.
\end{equation}
A suitable combination of the Pedersen and Hall diffusivities can lead to
$\kc \to \infty$. This occurs when the denominator in
Equation (\ref{eq:kc}) vanishes
\begin{equation}
  1 + (4-q)\tetah + 2(2-q)(\tetah^2 + \tetap^2) = 0.
\end{equation}
The wavenumber at which the growth rate is maximum is
\begin{equation}
\label{eq:km}
  \tilde{k}_{\mrm{m}}^2 = \frac{-2\gammam^2[\gammam^2 + 2(2-q)]}
  {2\gammam^2 - 2q - [\gammam^2 + 2(2-q)]
  (q\tetah - 2\gammam\tetap)} ,
\end{equation}
and the maximum growth rate $\gammam$ normalized by $\Omega$ satisfies
\begin{equation}
\label{eq:gammam12}
  \tetah = \frac{16q\tetap\gammam}{4q^2- 16\gammam^2}
  - \frac{2}{\gammam^2 + 2(2-q)}.
\end{equation}

In a portion of the parameter space defined by $(\tetah, \tetap)$,
the maximum growth rate is reached asymptotically as the wavenumber approaches
infinity and the denominator of Equation (\ref{eq:km}) vanishes. The
growth rate in this region is obtained by solving
\begin{equation}
\label{eq:gammam3}
[\gammam^2 + 2(2-q)](\tetah^2 + \tetap^2)
  + [2\tetap\gammam + (4-q)\tetah] + 1= 0 \, .
\end{equation}
This regime will be the subject of greater discussion in the following
section.

\begin{figure}
  \centering
  \includegraphics{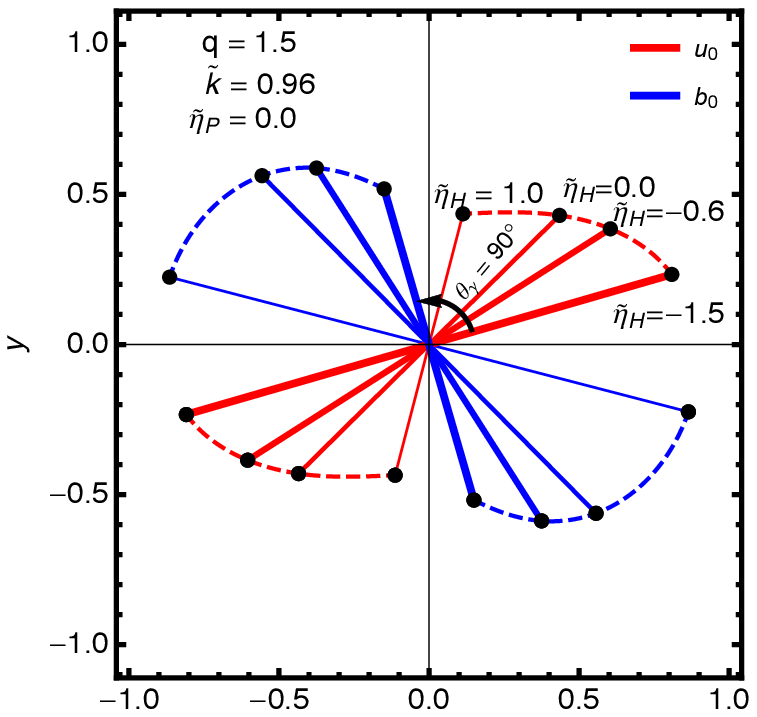} \\ \bigskip
  \includegraphics{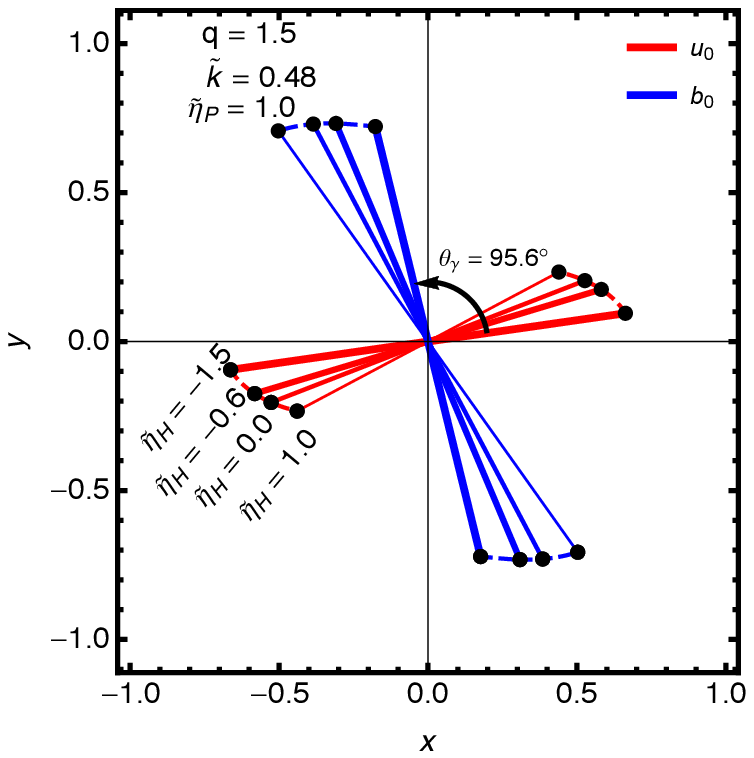}
  \caption{A geometrical representation of the velocity field $\bs
  {\delta_b}$ and the magnetic field
  $\bs{\delta_b}$ projected on the horizontal plane of the disk for different
  values of the Hall parameter, $\tetah = -1.5, -0.6, 0.0, 1.0$. The top panel
  presents the case without dissipation $\tetap = 0$ evaluated at
  $\tilde{k} = 0.96$ and the bottom panel illustrates the case with
  magnetic dissipation $\tetap = 1.0$ evaluated at the wavenumber
  $\tilde{k} = 0.48$. A general trend that one observes is for the
  velocity vector to lean in towards the positive $x$ axis and for the
  magnetic vector to lean in towards the positive $y$ axis with increasingly
  negative Hall parameter. The magnetic and velocity vector are however only
  orthogonal to each other in the dissipationless limit and when $\Pm = 0$.
  }
  \label{fig:geogamma}
\end{figure}
Let us now examine how the planes containing the velocity and magnetic vectors
$\bs{\delta_u}$ and $\bs{\delta_b}$ associated with the unstable mode are
oriented relative to each other.
Using Equation (\ref{eq:evrelationship}), we find
\begin{align}
  \cos{\theta_{\gamma}} \equiv
  \frac{\eg^1 \eg^3 + \eg^2 \eg^4}
  {b_0 u_0}
  = -\frac{\omga \omgp q\Omega \eg^1}
  {b_0 u_0} \, .
  \,
\end{align}
In the absence of dissipation, $\omgp \to 0$, $\theta_{\gamma} =
\pi/2,3\pi/2$, and $\bs{\delta_u}$ and $\bs{\delta_b}$ are
orthogonal to each other.
Additionally, the angle $\psi$ subtended by the
velocity vector $\bs{\delta_u}$ with respect to the $x$ axis in the $xy$ plane
is simply given by $\tan\psi = |\eg^2|/|\eg^1|$.
\begin{figure}
  \centering
  \includegraphics{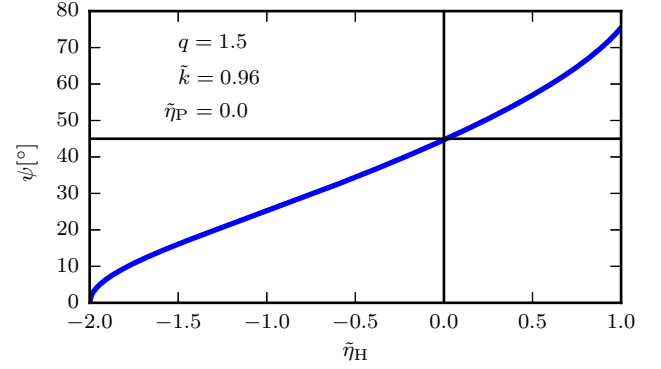}
  \caption{ The angle $\psi$ between the velocity vector $\bs{\delta}_u$
  and the $x$ axis plotted as a function of the Hall
  parameter $\tetah$. The angle is calculated from the eigenvector components
  evaluated at the wavenumber, $\km$ for the ideal MRI. }
  \label{fig:psi}
\end{figure}
\begin{figure*}
  \centering
  \includegraphics{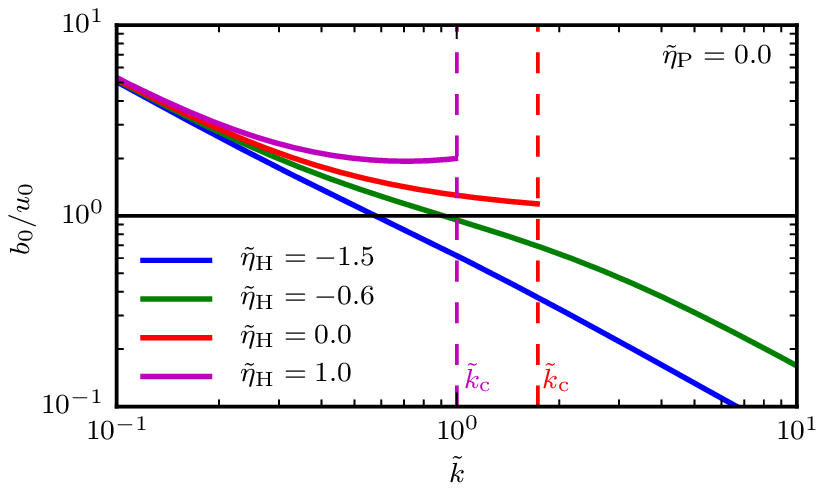}
  \includegraphics{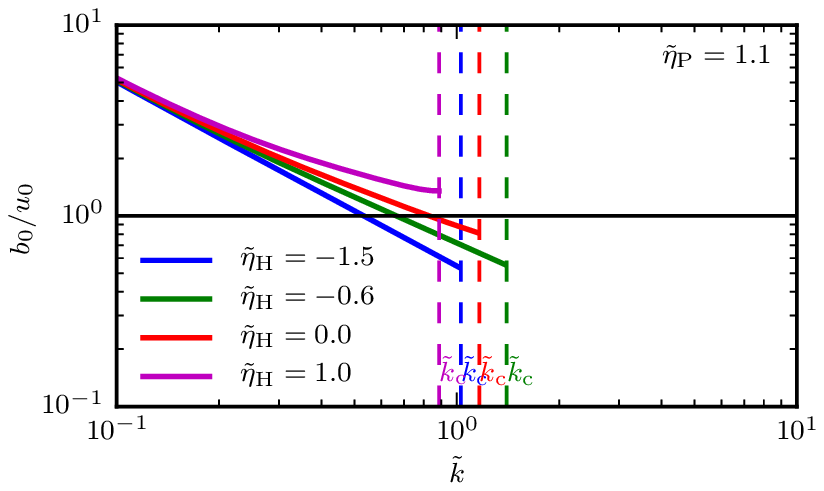}
  \caption{The ratio $b_0/u_0$ of the MRI unstable eigenmode for
  different values of the Hall parameter $\tetah$ without dissipation (left)
  and with dissipation (right).
  This ratio becomes lesser than unity
  for $\tetah < 0$ and it has implications for the relative strengths of the
  magnetic and kinetic stresses pertinent to angular momentum transport.}
  \label{fig:bvu}
\end{figure*}

Figure \ref{fig:geogamma} illustrates $\bs{\delta_u}$ and $\bs{\delta_b}$
projected on to the mid-plane of the disk for four representative values of
the Hall diffusivity $\tetah$, for a fixed wavenumber $\tilde{k}$, with and
without
dissipation $\omgp$.
The angle $\psi$ becomes smaller with increasingly negative values
of the Hall parameter, $\tetah$.
This is shown graphically in Figure \ref{fig:psi} for the wavenumber
$\km$ at which the growth rate of the ideal MRI is maximum.
One can also see that the velocity and magnetic vectors are not
quite orthogonal when $\omgp \neq 0$ \citep{pc08}.

Finally, the ratio of the magnitudes of the magnetic vector to the velocity
vector, $b_0/u_0$, can also be computed from the eigenvector components
Equation (\ref{eq:fulleigvector}).
Figure \ref{fig:bvu} measures this ratio
as a function of wavenumber for different values
of the Hall parameter.
We find that this ratio becomes lesser than unity implying that the
magnetic perturbation is weaker in comparison to the velocity perturbation
when $\tetah < 0$ and for a very large range of wavenumbers with
$\tetap < 1$.
This feature will be of particular interest with regard to
the transport stresses of the non-ideal MRI unstable mode
and will be explored further in the following section.

\section{Kinetic and Magnetic Stresses and Energy Densities}
\label{sec:energetics}

We now use the results of the eigenmode analysis to
ascertain the properties of the mean kinetic and magnetic
stresses and energy densities. In particular, we focus on the
$xy$ component of the Reynolds and Maxwell stresses of the
MRI mode.
We define the mean Reynolds and Maxwell stresses as
\begin{equation}
\label{eq:stress-def}
  R_{ij}(t) = \overline{\delta u_i(z,t) \delta u_j(z,t)}
  \,\: \text{and} \,\;
  M_{ij}(t) = \overline{\delta b_i(z,t) \delta b_j(z,t)},
\end{equation}
where the over-line denotes the vertical
average over the domain $[-H, H]$. In terms of their Fourier
components, the stress components are given by
(see \citealt{pcp06a} for the derivation)
\footnote{In order to keep track of the various modes contributing
to the mean values, we restore the wavenumber index $n$ throughout this
section.}
\begin{align}
\label{eq:reynolds-stress}
  R_{ij}(t) &\equiv 2\sum_{n=1}^{\infty}
  \Re[\hat{\delta u_i}(\kn,t)\hat{\delta u_j^*}(\kn,t)],
  \\
\label{eq:maxwell-stress}
  M_{ij}(t) &\equiv 2\sum_{n=1}^{\infty}
  \Re[\hat{\delta b_i}(\kn,t)\hat{\delta b_j^*}(\kn,t)].
\end{align}
The $xy$ component of the Reynolds and Maxwell stress tensor associated with
the Hall-MRI unstable eigenmode are
\begin{align}
  \label{eq:reynolds-xyt}
  R_{xy}(t) &=
  2\sum_{n=1}^{\infty} \mathcal{R}_{xy}(\kn)\mrm{e}^{2\sigma(\kn) t},
  \\
  \label{eq:maxwells-xyt}
  M_{xy}(t) &=
  2\sum_{n=1}^{\infty} \mathcal{M}_{xy}(\kn)\mrm{e}^{2\sigma(\kn) t},
\end{align}
where
\begin{align}
  \label{eq:reynolds-xyk}
  \mathcal{R}_{xy}(\kn) &=
  \frac{\Re[e_{\gamma}^1 e_{\gamma}^{2*}]}{||\bs{e}_{\gamma}||^2},  \\
  \label{eq:maxwells-xyk}
  \mathcal{M}_{xy}(\kn) &=
  \frac{\Re[e_{\gamma}^3 e_{\gamma}^{4*}]}{||\bs{e}_{\gamma}||^2}.
\end{align}

The trace of the tensors $R_{ij}$ and $M_{ij}$ gives us the mean kinetic and
magnetic energy densities respectively
\begin{align}
\label{eq:kinergy-t}
  E_{\mrm{K}}(t) &=
  2\sum_{n=1}^{\infty} \mathcal{E}_{\mrm{K}}(\kn)\mrm{e}^{2\sigma(\kn) t},
  \\
  \label{eq:manergy-t}
  E_{\mrm{M}}(t) &=
  2\sum_{n=1}^{\infty} \mathcal{E}_{\mrm{M}}(\kn)\mrm{e}^{2\sigma(\kn) t},
\end{align}
where
\begin{align}
  \label{eq:kinergy-k}
  \mathcal{E}_{\mrm{K}}(\kn) &=
  \frac{\mathcal{R}_{xx}(\kn) + \mathcal{R}_{yy}(\kn)}{2},  \\
  \label{eq:manergy-k}
  \mathcal{E}_{\mrm{M}}(\kn) &=
  \frac{\mathcal{M}_{xx}(\kn) + \mathcal{M}_{yy}(\kn)}{2}.
\end{align}
The quantities $\mathcal{R}_{xy}$ , $\mathcal{M}_{xy}$,
$\mathcal{E}_{\mrm{K}}$ and $\mathcal{E}_{\mrm{M}}$ represent the contribution
of each mode $k$ to the mean values of the corresponding functions
\citep{pcp06a}.

The ratio of the $xy$ components of Maxwell stress to the Reynolds stress
is a non-trivial function of $\kn$.
In the ideal limit (with $\omgv = \omgp = \omgh = 0$), using the dispersion
relation, one can easily see that $\mathcal{M}_{xy} > \mathcal{R}_{xy}$ for
the full range of unstable modes, $\kn$.
In the dissipationless limit, where $\omgv = \omgp = 0$ but $\omgh \neq 0$,
this ratio reduces to
\begin{equation}
\label{eq:stressratios}
  \frac{-\mathcal{M}_{xy}(\kn)}{\mathcal{R}_{xy}(\kn)} =
  \frac{\omga^2(2\Omega - \omgh)^2}{[\gamma^2(\kn) + \omga^2 +
  \kappah^2]^2}.
\end{equation}
Interestingly,  the ratio defined in Equation (\ref{eq:stressratios}) is only greater
than unity if
\begin{equation}
\label{eq:ki}
  \tilde{k}_{\mrm{n}}^2  < \tilde{k}_{\mrm{i}}^2  = \frac{(q-2)}{\tetah} \,.
\end{equation}
The wavenumber $\ki$ is purely imaginary if $\tetah>0$ and infinite valued if $\tetah=0$.
However, when $\tetah<0$ and $q<2$, $\ki$ is finite and real valued.
This implies that there is a range of unstable wavenumbers for which
$\mathcal{R}_{xy} > \mathcal{M}_{xy}$. It is rather difficult to derive
an equivalent expression for $\ki$ in closed form with $\omgp \neq 0$
since this would require solving a quartic equation in both $\tilde{k}$
and $\gamma$.
However, numerical calculations hint at the presence of such a scale
with dissipative effects present as well and we comment further on this in
the following section. As we shall discuss below, the
potential for a role-reversal of the dominant stress components are
directly tied to the exact nature of the unstable mode in different parts of
parameter space.

The characteristic variables that specify the wavenumber at
which the growth rate is quenched $\kc$, and the wavenumber at which the
growth rate is maximum $\km$, divides the parameter space defined by $
(\tetah,\tetap)$ into three regions $\mathrm{I}$, $\mathrm{II}$ and $\mathrm
{III}$ as described in \cite{ws12}. Region $\mathrm{I}$ is defined by
the space outside of a semi-circle in the coordinates $(\tetah,\tetap)$ spanning
from $(-1/2,0)$ to $(-2,0)$. Here the unstable mode has a finite $\kc$ and
$\km$. The space contained within the aforementioned semi-circular locus and
an arc extending from $(\tetah,\tetap) = (-4/5,0)$ to $(-2,0)$ is designated
Region $\mathrm{II}$. Here the unstable mode has a finite $\km$ but $\kc$ is
infinite. Finally, the area enclosing the lower boundary of Region $\mathrm
{II}$ and the horizontal axis $\tetah$ is designated Region $\mathrm{III}$.
In this region, both $\kc$ and $\km$ are infinite.
The region $\tetah<-2$ is stable to the MRI for all values of $\tetap$.
This classification will be useful in specifying the dominant stresses in
parameter space as we discuss below.

\subsection{Stresses and Energies in Region $\mathrm{I}$}
\label{sec:stregion1}

As mentioned above, the MRI growth is cut-off
at a finite wavenumber in Region $\mrm{I}$. This implies that the
major contributions to Equations (\ref{eq:reynolds-xyt}),
(\ref{eq:maxwells-xyt}), (\ref{eq:kinergy-t}) and (\ref{eq:manergy-t})
come from a finite range of unstable wavenumbers $n=1$ to $n= N_{\mrm{c}}$
where $N_c$ labels the cut-off wavenumber $\kc$.
At late times, the mean stresses and energy densities may
then be expressed as
\begin{align}
  \label{eq:reynolds-xyt-etap}
  R_{xy}(t) &=
  2\sum_{n=0}^{N_{\mrm{c}}} \mathcal{R}_{xy}(k_n)\mrm{e}^{2\gamma(k_n) t} +
  \dots,
  \\
  \label{eq:maxwells-xyt-etap}
  M_{xy}(t) &=
  2\sum_{n=0}^{N_{\mrm{c}}} \mathcal{M}_{xy}(k_n)\mrm{e}^{2\gamma(k_n) t} +
  \dots,
  \\
  E_{\mrm{K}}(t) &=
  2\sum_{n=0}^{N_{\mrm{c}}} \mathcal{E}_{\mrm{K}}(k_n)\mrm{e}^{2\gamma(k_n)
  t} +
  \dots,
  \\
  E_{\mrm{M}}(t) &=
  2\sum_{n=0}^{N_{\mrm{c}}} \mathcal{E}_{\mrm{M}}(k_n)\mrm{e}^{2\gamma(k_n)
  t} +
  \dots \,,
\end{align}
with the dots representing oscillatory
contributions that we may safely neglect.
Within this region of parameter space, it is reasonable to
expect that at late times during the linear evolution, the kinetic and
magnetic stresses are dominated by contributions linked to the scale $\km$.
In the dissipationless limit, we can thus expect
\begin{equation}
\label{eq:stressposeta}
  \lim_{t\Omega \gg 1} \frac{-M_{xy}}{R_{xy}} \sim
  \left. \frac{-\mathcal{M}_{xy}}{\mathcal{R}_{xy}} \right|_{\km}
  = \frac{(4 - q)[(4 - q)\tetah + 2]}{2q}.
\end{equation}
Equation (\ref{eq:stressposeta}) trivially reduces to Equation (65) of \citet{pcp06a} in the ideal MHD limit.
Deriving an equivalent analytical expression for the late time stress
ratios in the presence of dissipation is tedious but can easily be computed
numerically. However, numerical calculations also reveal
that a real valued $\ki$ may be present for certain values of $\tetap$
in Region $\mathrm{I}$
and the scales are arranged in the order $\km < \ki \lesssim \kc$.
Nevertheless, the ratio of the stress components will be dominated by
the fastest growing mode, at which one always finds $-\mathcal{M}_{xy} >
\mathcal{R}_{xy}$. In the dissipationless limit, $\ki$ is never real valued
in Region $\mathrm{I}$.

\begin{figure}
  \centering
  \includegraphics[width=8.5cm,height=10cm,keepaspectratio]{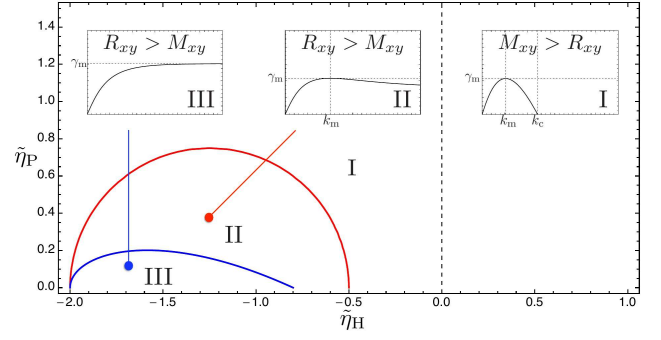}
  \caption{The parameter space defined by $\tetah$ and $\tetap$ demarcated into
  three regions $\mathrm{I},\mathrm{II}$ and $\mathrm{III}$ based on the
  distinct characteristic traits of the MRI for the said range of parameter
  values.
  The figure is identical to Fig 5 of \cite{ws12} with the
  relative strengths of the $xy$ kinetic and magnetic stress components
  additionally specified.
  }
  \label{fig:paraspace}
\end{figure}

\subsection{Stresses and Energies in Regions $\mathrm{II}$ and $\mathrm{III}$}
\label{sec:stregion23}

The unstable mode grows at a uniform rate for a wide range of wavenumbers that
extend infinitely in both Regions $\mathrm{II}$ and $\mathrm{III}$. One can
therefore derive asymptotic forms of the
\textit{per-k} kinetic and magnetic stress energy densities,
Equations (\ref{eq:reynolds-xyk}),
(\ref{eq:maxwells-xyk}), (\ref{eq:kinergy-k}) and (\ref{eq:manergy-k})
as given below

\begin{figure*}
  \centering
  \includegraphics{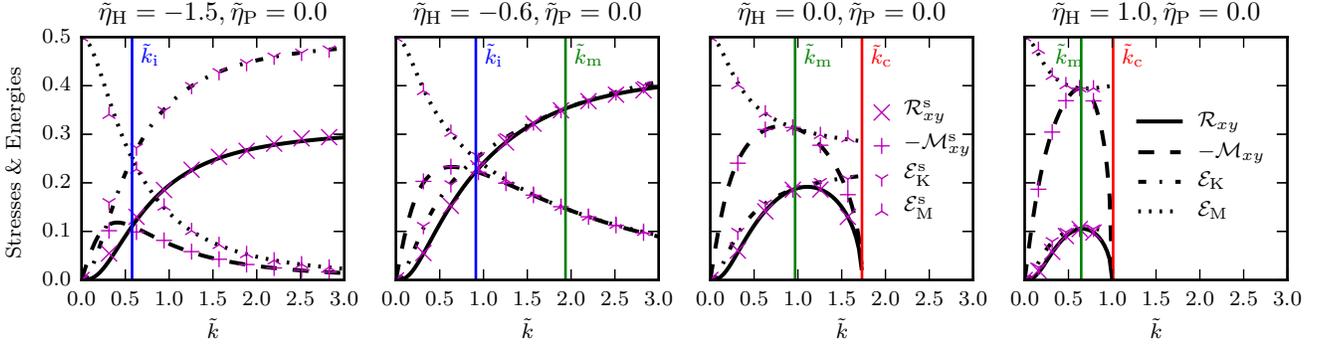}
  \caption{The $xy$ components of the \textit{per-k} Reynolds and Maxwell's
  stress tensor and the kinetic and magnetic energy densities of the MRI
  unstable mode,
  Equations (\ref{eq:reynolds-xyk}),
  (\ref{eq:maxwells-xyk}), (\ref{eq:kinergy-k}) and (\ref{eq:manergy-k}),
  for different values of the Hall parameter, $\tetah = -1.5, -0.6, 0.0,
  1.0$ and $q = 1.5$ are plotted by the line curves. The discrete markers
  denote the corresponding values of the said quantities derived from
  shearing box simulations. Legends with the superscript `$s$'
  label the corresponding quantity derived from simulation data.}
  \label{fig:hevp-comstresses}
\end{figure*}
\begin{figure*}
  \centering
  \includegraphics{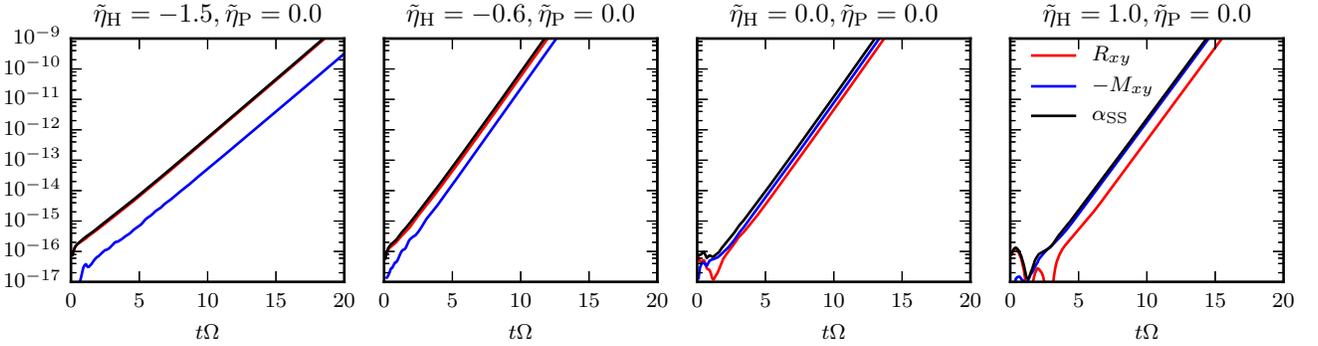}
  \caption{The $xy$ components of the Reynolds stress tensor, Maxwell's
  stress tensor and the Shakura-Sunyaev $\alpha_{\mrm{SS}}$ parameter of the
  MRI unstable mode for different values of the Hall parameter, $\tetah =
  -1.5, -0.6, 0.0,
  1.0$ and $q = 1.5$ obtained from shearing box simulations with $N_z = 256$.
  In accordance with the results of the linear theory, we find that the
  Reynolds stress dominates over the Maxwell's stress when $\tetah < 0$. }
  \label{fig:rowalpha}
\end{figure*}

\begin{align}
\label{eq:reynolds-xyk-asym}
  \lim_{\tilde{k} \to \infty} \mathcal{R}_{xy} &\sim
  \frac{\etat^2(2\Omega+\ua^2 \etah/\etat^2)(\gamma_{\infty} +
  \ua^2\etap/\etat^2)}{\ua^4 + 2\ua^2(\gamma_{\infty}\etap + 2\Omega\etah)
  + \etat^2(\gamma_{\infty}^2 +4\Omega^2)},
  \\
\label{eq:maxwells-xyk-asym}
  \lim_{\tilde{k} \to \infty} \mathcal{M}_{xy} &\sim 0,  \\
\label{eq:kinergy-xyk-asym}
  \lim_{\tilde{k} \to \infty} \mathcal{E}_{\mrm{K}} &\sim \frac{1}{2}, \\
\label{eq:manergy-xyk-asym}
  \lim_{\tilde{k} \to \infty} \mathcal{E}_{\mrm{M}} &\sim 0,
\end{align}
where $\etat^2 = \etah^2 + \etap^2$ and $\gamma_{\infty}$ is the solution to
Equation (\ref{eq:gammam12}) for Region $\mathrm{II}$ and
Equation (\ref{eq:gammam3}) for Region $\mathrm{III}$.
Using Equations (\ref{eq:reynolds-xyk-asym}) and (\ref{eq:kinergy-xyk-asym})
in Equations (\ref{eq:reynolds-xyt}) and (\ref{eq:kinergy-t}),
we may then approximate the time dependent $xy$ Reynolds stress tensor and
kinetic energy density as
\begin{align}
  \label{eq:reynolds-xyt-asym}
  R_{xy}(t) &\approx
  2 \mrm{e}^{2\gamma_{\infty} t} \mathcal{R}_{xy}(\tilde{k} \to \infty)
  \sum_n^{\infty} 1^n,
  \\
  \label{eq:kinergy-t-asym}
  E_{\mrm{K}}(t) &\approx \mrm{e}^{2\gamma_{\infty} t}
  \sum_n^{\infty} 1^n.
\end{align}

While the infinite sum in Equations (\ref{eq:reynolds-xyt-asym}) and
(\ref{eq:kinergy-t-asym}) appear to be a divergent
series, it is in fact the Riemann zeta function
\[
\zeta(s) = \sum_{n=1}^{\infty} \frac{1}{n^s} \, ,
\]
with $s=0$ and possesses a finite sum $\zeta(0) = -1/2$ \citep{hardy}.
We shall not endeavour to speculate on the implications of this curious
feature since an infinite range of scales will never come to pass as
the fluid approximation inevitably breaks down. The alternative is no less
dramatic in that a finite series would have the sum $ \sum_{n=1}^N 1^n =N$
where $N$
can be rather large.

We are thus led to expect
\begin{equation}
  \lim_{t\Omega \gg 1}  \frac{-M_{xy}}{R_{xy}} \ll 1 \, ,
\end{equation}
with the ratio becoming increasingly smaller the greater the unstable range of
wavelengths accounted for.
In a real astrophysical system such as a protoplanetary disk,
dissipation due to ohmic and ambipolar
diffusion may be large enough in some parts of the
disk to keep the kinetic stress $R_{xy}$ and energy density
$E_{\mathrm{K}}$, bounded, by suppressing the unstable growth at
smaller length scales. Therefore, the dominance of kinetic stresses may go
unchallenged unless
dissipation forces the instability to operate within Region $\mathrm{I}$,
see Figure \ref{fig:paraspace}. On the other hand, if one can find parts of
the disk where the diffusivities fall within Regions $\mathrm{II}$ and
$\mathrm{III}$, one should expect
the Reynolds stress to dominate. Figure \ref{fig:hevp-comstresses} shows the
\textit{per-k} kinetic and magnetic $xy$ stress component and energy densities
in the dissipationless limit for different values of the Hall parameter,
$\tetah$.

\section{Comparing Analytical results with Numerical Simulations}
\label{sec:simulations}

In this section we present the results of unstratified shearing box
simulations with a uniform net vertical field including Hall and diffusion, performed using
the grid-based higher order Godunov MHD code ATHENA \citep{athena08}.
The Hall effect is implemented in Athena using an operator-split technique
\citep{bai2014}
that is similar to the dimensionally split scheme proposed by
\citet{sd2006,sd2007}.
We use the HLLD Riemann solver and a CTU unsplit integrator with
third order reconstruction.
The simulations we performed are identical to the test runs reported in
Appendix B of \citealt{bai2014}.

We adopt an isothermal equation of state and the initial conditions constitute
random velocity perturbations of strength, $\delta u/c_{\mrm{s}} = 10^{-6}$.
The default
boundary conditions are periodic in $y$ and $z$ and shearing periodic in $x$.
Our simulations
were performed with a plasma beta, defined as the ratio of thermal to magnetic
pressure $\beta = 800$, background angular frequency $\Omega = 1$,
equilibrium density $\rho_0 = 1$, isothermal sound speed $c_{\mathrm{s}} =
1$ and dimensionless shear rate $q = 3/2$.
The computational domain has an extent of $L_x \times L_y \times L_z = 0.1H
\times 0.1H \times 2H$. We work with the default grid resolution
$N_x \times N_y \times N_z = 4 \times 4 \times 256$.

In order to directly test and
compare against the predictions of analytical theory, we run the code
by varying the Hall parameter over the different values, $\tetah =
-1.5,-0.6,0.0,1.0$ and $\tetap = 0.0$.\footnote{Note that the dimensionless
Hall
parameter in Athena, $Q_H$, is
related to the Hall parameter in our work as $Q_H = \sqrt{2/\beta}\,\tetah$.}
We also perform one additional simulation with the parameters $\tetah =
-1.5$ and $\tetap = 1.0$.
The simulations were run for up to $20$ orbits with orbital advection via Fargo
enabled. Such short run times suffice for the task at hand since the aim is to test
the agreement between our analytical results and the linear evolution of the simulations.
We obtain the perturbations, $\dux, \duy, \dbx, \dby$, from the Athena output
and compute their Fourier transform at time, $t = 11
\Omega^{-1}$. We then combine these variables as given by
Equations (\ref{eq:reynolds-xyk}), (\ref{eq:maxwells-xyk}),
(\ref{eq:kinergy-k}) and (\ref{eq:manergy-k})
to obtain the kinetic and magnetic stress components and energies at a given scale.

\begin{figure}
  \centering
  \includegraphics{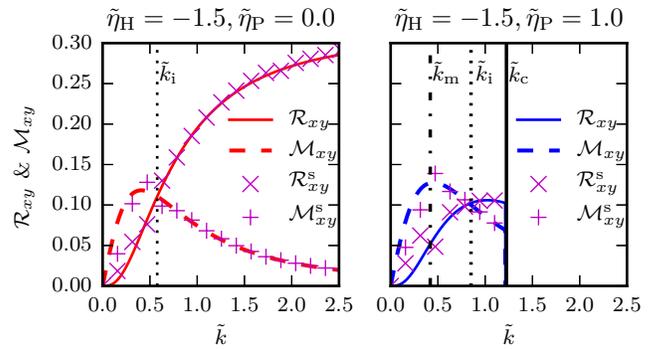}
  \caption{The Reynolds and Maxwells stress component $\mathcal{R}_{xy}$ and
  $\mathcal{M}_{xy}$ of the MRI unstable mode
  with $\tetah = -1.5$ and $q = 1.5$  without dissipation
 $\tetap = 0$ (left panel) and with dissipation $\tetap = 1.0$ (right panel).
  The discrete markers denote the corresponding
  values of the stresses derived from shearing box simulations.
  Legends with the superscript `$s$'
  label the corresponding quantity derived from simulation data.
  }
  \label{fig:stressesped}
\end{figure}

We have found the simulation and the theoretical results to be in excellent
agreement for as many vertical modes, $\tilde{k}$, as can be reliably resolved.
The output of the shearing box simulation conducted with a vertical grid
resolution, $N_z = 256$, is over-plotted against the
values of the corresponding stresses and energy densities obtained from linear
theory in Figure \ref{fig:hevp-comstresses}.
Figure \ref{fig:rowalpha} plots the growth in the $xy$ time dependent Reynolds
and Maxwell's stress as well as the Shakura-Sunyaev alpha parameter defined as
\begin{equation}
  \alpha_{\mrm{SS}} \equiv \frac{\int (\overline{\rho \dux \duy} -
  \overline{\dbx \dby})dz}{c_{\mathrm{s}}^2 \int \rho dz},
\end{equation}
for the same set of parameters $\tetah = -1.5,-0.6,0.0,1.0$, $\tetap = 0.0$
and $N_z = 256$ and where the overlines denote horizontal averages. In
accordance with the implications that followed from
Equations (\ref{eq:reynolds-xyt-asym}) and (\ref{eq:kinergy-t-asym}), we find
that even for such
moderate resolutions, the Reynolds stress noticeably dominates the Maxwell's
stress during the linear growth of the instability.
For a fixed value of $\tetah = -1.5$, we compare the kinetic and magnetic
stress and energy densities with two different values of $\tetap = 0.0,1.0$
in Figure \ref{fig:stressesped}. Although a finite value of $\ki$ appears to
be present with $\tetap=1.0$, $-\mathcal{M}_{xy} \gg \mathcal{R}_{xy}$ at
$\km$ and so Maxwells stress maintains its hegemony over its kinetic
counterpart.

\begin{figure}
  \centering
  \includegraphics{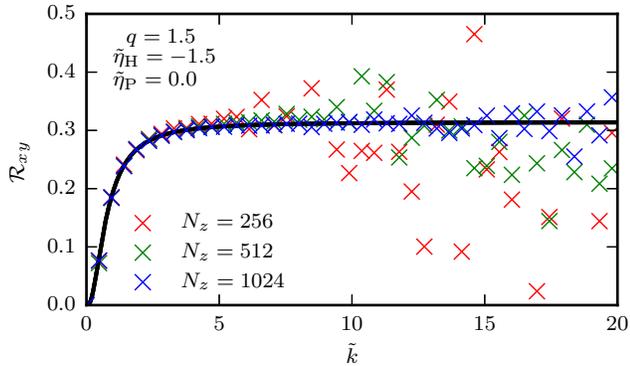}
  \caption{The \textit{per-k} Reynolds stress component $\mathcal{R}_{xy}$ of
  the MRI unstable mode
  with $\tetah = -1.5$ and $q = 1.5$. The crosses denote the corresponding
  values of the \textit{per-k} stress component derived from shearing box
  simulations with three
  different vertical grid resolutions, $N_z = 256, 512, 1024$.
  The agreement between analytical and
  numerical results improves at higher wavenumbers as the resolution
  increases.
  }
  \label{fig:resolution}
\end{figure}

Figure \ref{fig:resolution} compares the values of the $xy$ component of
the \textit{per-k} Reynolds stress tensor obtained from simulations with three
different vertical grid resolutions. It is quite apparent that with
increasing resolution, the agreement between theory and simulation improves
substantially as many more smaller scale modes are reliably resolved.
This places a stringent requirement upon the resolution
demands while performing simulations of a weakly magnetized shearing system
when Hall diffusion is present and dissipation is comparatively weak, if one
is to obtain accurate results in accordance with theoretical expectations.
In the simulations conducted by \citet{ss02b}, the vertical resolution was
generally low $(N_z = 32, L_z = H)$. However, one can already see in their
results that the volume averaged Reynolds and Maxwell's stresses at saturation
were the same order of magnitude when $\tetah < 0$ and $\tetap < 1$.
This is not so for comparable simulations performed with resistivity but
without Hall diffusion \citep{sits2004} where the $xy$ Maxwell's stress at
saturation was larger than the corresponding Reynolds stress.
While we have not explored the non-linear regime in our work, we anticipate
that with higher grid resolution, one might find stronger mean Reynolds
stress perpetuating even at late times. This could be confirmed with dedicated
numerical studies.


\section{Summary and Discussion}
\label{sec:discussion}

In this paper, we have carried out a detailed examination
of the linear eigenmodes in the shearing sheet framework for a weakly
magnetized system subject to non-ideal effects with special focus on
Hall diffusion.
Although our analysis invoked simplifying assumptions, we have nonetheless
been able to go a step further from similar analysis performed in the past and
glean certain key attributes governing these modes.
A careful examination of the eigenvectors has enabled us to provide
a detailed description of the polarization properties and to sketch a
visual representation of the eigenmodes as they evolves in space and time.
By employing the formalism of \citet{pcp06a},
we have also derived expressions for the kinetic and magnetic stresses and
energy densities in terms of the complex eigenvector components.
This has enabled us to generalize the ratio of the magnetic to kinetic
stresses applicable to the later stages of linear evolution of the MRI when
subject to Hall diffusion.
Our central result is the identification of regimes in the parameter space
defined by $(\tetah,\tetap)$ wherein the kinetic stresses and energies are
found to dominate their magnetic equivalents. This property is in sharp
contrast to what one expects of the ideal MRI or the MRI subject to
dissipative effects alone.

Since the non-ideal MRI unstable eigenmodes studied here are also exact
non-linear solutions of the shearing sheet equations \citep{kl13,goxu94},
the unique traits associated with these modes may carry through or influence
the subsequent non-linear evolution of the system.
In ideal as well as dissipative MHD \citep{pg09,lb09,p10},
these so-called channel modes have been shown to be unstable to parasitic
instabilities which may result in their ultimate saturation.
\citet{kl13} is the only work we are aware of that has
explored the stability of the Hall-MRI modes to parasitic instabilities.
In light of the findings presented here, it would be worthwhile to
revisit the question of saturation via parasitic modes, particularly for the case
with negative Hall diffusivities $(\tetah < 0)$ and weak dissipation.

There have been a number of recent numerical studies of a
weakly magnetized system subject to
Hall diffusion (\citealt{kl13,lesuretal2014,bai2014,bai2015,simon15}) in the
shearing box framework.
To our knowledge, none of these studies have reported anything resembling
the behavior of stresses with $\tetah <0$, that we have presented in this
paper. We surmise that this may be due to the
insufficient vertical grid resolution and comparably strong ohmic and
ambipolar diffusion present in virtually all of these simulations.
Most of these studies have been performed with primary applications to
protoplanetary disks and amongst them, simulations exploring the system with
anti-parallel angular momentum and magnetic field vectors have been
comparatively few.
\citet{simon15} did however report the appearance of transient turbulent
bursts in their shearing box simulations with all non-ideal effects and
anti-parallel angular momentum and magnetic field vectors. However, they
attribute this behavior to a non-axisymmetric version of the Hall-Shear
instability (\citealt{rudkit05,kunz08}).

Conventional wisdom dictates that the ensuing turbulence in a
magnetorotationally unstable system is one that is dominated by magnetic
stresses and energies.
Astrophysical disks such as those around young stellar objects are
thought to harbor regions within them where Hall diffusion is the dominant
non-ideal effect \citep{bt01,kb04,w2007,ws12,bai2011,xb16}. These regions may also be subject to
diffusion by ohmic and ambipolar diffusion to varying extents.
If the dissipative effects are
sufficiently strong, they can act to cut down the range of scales
unstable to the MRI and
thereby curtail the dominance of kinetic stresses if
$\tetah < 0$. However, there is no definitive estimate at the moment of how
prevalent the different non-ideal effects are and to what degree. Therefore,
it is still too early to judge whether factors that favor the conditions
leading to predominant kinetic stresses may or may not be found.
The implications that this role-reversal might have upon the ensuing
turbulence warrants further study.

\acknowledgements
We are grateful to the referee whose comments led to an improved version of
the paper.
We acknowledge useful discussions with Tobias Heinemann, Oliver Gressel and
Leonardo Krapp. We are grateful
to Thomas Berlok for help with the simulations and for useful comments on the
manuscript.
The research leading to these results has received funding from the
European Research Council under the European Union’s Seventh Framework
Programme (FP/2007-2013) under ERC grant agreement 306614.

\appendix

\section{Classification of the Eigenmodes in the dissipationless limit}
\label{sec:evasymptotics}

\begin{figure*}
  \centering
  \includegraphics{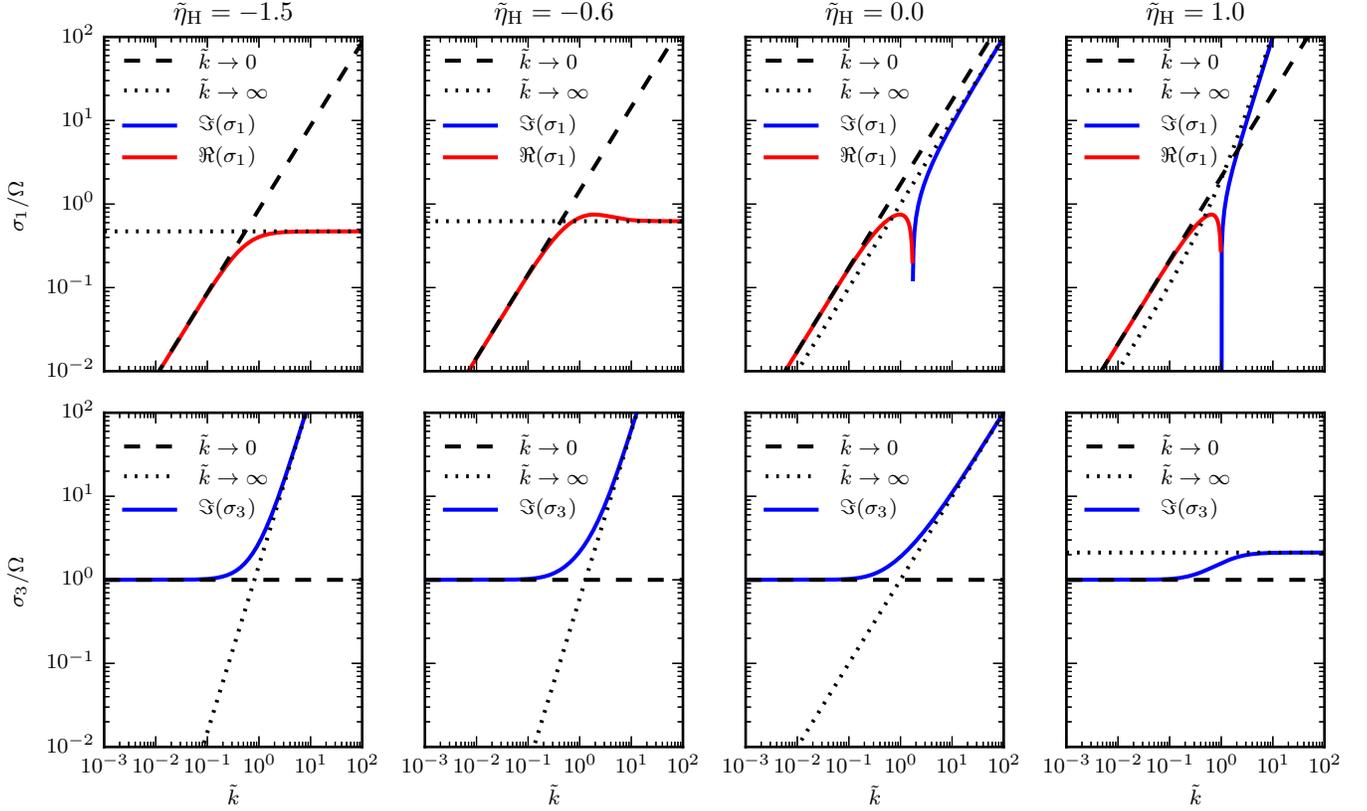}
  \caption{The positive branch solutions, $\sigma_1$ and $\sigma_3$, of the
  eigenvalue problem for four representative values of the Hall
  parameter,
  $\tetah = -1.5, -0.6, 0.0, 1.0$ and $q = 1.5$
  in the dissipationless limit. Solid lines represent
  the numerically computed eigenvalues (the real part in red and the imaginary
  part in blue). Asymptotic approximations in the low wavenumber limit
  ($\lim \tilde{k} \to 0$) and the high wavenumber limit
  ($\lim \tilde{k} \to \infty$) are
  represented by the dashed and dotted lines respectively.}
  \label{fig:hevp-eigvals}
\end{figure*}

Here, we solve the dispersion relation Equation (\ref{eq:dispersion}) in the
dissipationless limit $\omgv = \omgp = 0$ and describe the nature of
the different solutions in some detail.
In the limit $M \to 0$ and choosing the root such that $y = -L/2$ in
Equation (\ref{eq:yeqn}), we find that the roots of
Equation (\ref{eq:depressed}) given by Equation (\ref{eq:sol2}) reduces to
\begin{equation}
  \lim_{M\to 0}\sigma = \pm\sqrt{-\Lambda_0 \mp \sqrt{\Delta_0}},
\end{equation}
where
\begin{equation}
  \Lambda_0 = \frac{L_0}{2} \quad \text{ and } \quad
  \Delta_0 = \frac{L_0^2}{4} - N_0,
\end{equation}
and
\begin{align}
\label{eq:lambda0}
  \Lambda_0 &= \omga^2 + \frac{\kappa^2}{2} + \frac{\kappah^2}{2} , \\
\label{eq:delta0}
  \Delta_0 &= \left( \Omega - \frac{\omgh}{2} \right)^2
  \left[ \left(\omgh + \frac{\kappa^2}{2\Omega} \right)^2 + 4\omga^2 \right].
\end{align}
Setting $\omgh \to 0$ in Equations (\ref{eq:lambda0}) and (\ref{eq:delta0}),
we recover the ideal MRI solutions \citep{pcp06a}.
For the purpose of identification, we shall designate the four eigenvalues as
\begin{equation}
  \sigma_1 = \gamma, \quad \sigma_2 = -\gamma, \quad
  \sigma_3 = i\omega, \quad \sigma_4 = -i\omega,
\end{equation}
where
\begin{equation}
  \gamma = \sqrt{-\Lambda_0 + \sqrt{\Delta_0}} \quad
  \text{and} \quad
  \omega = \sqrt{\Lambda_0 + \sqrt{\Delta_0}} \,.
\end{equation}
The notation $\gamma$ and $\omega$ has been chosen to be redolent of the
unstable and oscillatory nature of the corresponding eigenmodes.
The positive branch eigensolutions, $\sigma_1$ and $\sigma_3$, have the
following asymptotic forms, at very low
and high wavenumbers
\begin{align}
\label{eq:asym-k0}
  \lim_{\tilde{k} \to 0}\sigma &\sim
  \begin{cases}
    \omga\sqrt{q(2-q)^{-1}+ q \tetah} \,\,
    \text{R.E.P if }\tetah \geq 0 \text{ and L.E.P if }\tetah < 0, \\
    i \kappa, \quad \text{L.E.P}
  \end{cases} \\
\label{eq:asym-kinf}
  \lim_{\tilde{k} \to \infty}\sigma &\sim
  \begin{cases}
  i \omgh , \quad \text{R.E.P if }\tetah \geq 0 \text{ and L.E.P if }\tetah <
  0, \\
  i \omgg , \quad \text{L.E.P}
  \end{cases}
\end{align}
where $\omgg$ is the so-called gyration frequency \citep{hq14}
\begin{equation}
\label{eq:epihall}
  \omgg =
  \sqrt{\left[ 2\Omega + \frac{\omga^2}{\omgh} \right]
  \left[ (2 - q)\Omega + \frac{\omga^2}{\omgh} \right]} \,.
\end{equation}
In the absence of rotation and shear, $\omgg$ corresponds to
the ion-cyclotron frequency,
$\omega_\mathrm{ci} = eB/m_{\mathrm{i}}c$
reduced by the ionization fraction $n_e/n$.
The acronyms R.E.P and L.E.P stand for Right and Left Elliptically
Polarized respectively and indicates the direction
of polarization of the oscillatory eigenmodes as seen by an observer looking
down perched above the disk midplane.

The Coriolis force and the Hall effect endow the oscillatory modes with
a circular polarization or helicity. The effect of shear is to make the
polarization elliptical. Hall diffusion has the added effect of bringing
about divergent behavior of the oscillatory modes at large wavenumbers. One
of the otherwise Alfv\'enic branches breaks out into what is commonly referred
to as the Whistler mode where the frequency varies quadratically with
wavenumber. The other Alfv\'en branch asymptotes to a maximum frequency
corresponding to the reduced ion-cyclotron frequency as the
wavelength grows smaller.

Under ideal MHD conditions, an infinitesimal perturbation executes a circular
trajectory due to the Coriolis force. The shear eccentrically stretches
this motion towards positive azimuth inwards from the point of origin and
towards negative azimuth outwards. The Lorentz tension is activated and tries
to restore the fluid element thereby transferring angular momentum from the
inward moving fluid element to the tethered element moving outwards. The respective fluid elements fall further inwards and outwards to compensate
and the egression is greater at intermediate lengthscales where tension is
weakest. This is the standard physical picture of the MRI \citep{bh98}.
When $\tetah > 0$, the Hall effect introduces an ``epicyclic motion" of its
own \citep{bt01} that has the opposite sense of the Coriolis induced
epicycles.
At smaller length scales, this push-back is intensified and together with
tension, suppresses any unstable motion. When $\tetah < 0$, the Hall effect
induced epicycles have the same sense as the Coriolis motion and moreover
acts to negate the restoring magnetic tension forces at the smaller
lengthscales. These epicycles respond at the frequency $\omgg$ which is also
now purely imaginary and leads to continued exponential growth at ever smaller
lengthscales. \citet{ws12} refer to the instability as operating in the
``cyclotron limit'' at the high wavenumber end.

Figure \ref{fig:hevp-eigvals} shows the positive eigensolutions,
$\sigma_1$
and $\sigma_3$ as a function of wavenumber for four representative values
of $\tetah$. The asymptotic forms given by Equations (\ref{eq:asym-k0}) and
(\ref{eq:asym-kinf}) are plotted over the exact solutions for comparison.
Notice the eigensolutions $\sigma_1$ and $\sigma_3$, splitting into separate
branches with $\tetah = 1$ in Figure \ref{fig:hevp-eigvals},
at high wavenumbers.
For the sake of identification, we shall
refer to modes that asymptote to the frequency $\omgg$, as simply the
cyclotron
mode. Bear in mind however that when $-1/2<\tetah<\infty$,
$\sigma_1$ becomes oscillatory beyond the cut-off wavenumber $\tilde{k}_c$.
The change in sign of $\tetah$ effects an interchange of the Whistler and
cyclotron behavior on the modes, $\sigma_1$ and $\sigma_3$, at high
wavenumbers. Furthermore
when $-2<\tetah<-1/2$, $\omgg$ is purely imaginary and corresponds to
the large wavenumber growth rate of the unstable mode, $\sigma_1$.


\end{document}